\pdfoutput=1

\documentclass[11pt]{article}

\usepackage[]{acl}

\usepackage{times}
\usepackage{latexsym}
\usepackage{graphicx}
\usepackage{amsmath}
\usepackage{amssymb}
\usepackage{spverbatim}
\usepackage{verbatimbox}
\usepackage{xcolor}
\usepackage{newverbs}
\usepackage{cprotect}
\usepackage{fancyvrb}
\usepackage{float}

\definecolor{ao(english)}{rgb}{0.0, 0.5, 0.0}
\newverbcommand{\cverb}{\bf \color{blue}}{}
\newverbcommand{\gverb}{\bf \color{green}}{}

\newcommand{\block}[1]{%
  \raisebox{\dimexpr(\fontcharht\font`X-1em)/2}{\rule{0.5em}{#1\dimexpr1em/8}}%
}
\DeclareUnicodeCharacter{2581}{\block{1}}

\usepackage[T1]{fontenc}

\usepackage[utf8]{inputenc}

\usepackage{microtype}

\usepackage{inconsolata}

\let\oldfootnote\footnote
\def\footnote{\ifhmode\unskip\fi\oldfootnote}

\title{Probing Pretrained Models of Source Codes}

\author{Sergey Troshin, Nadezhda Chirkova\thanks{Now at Naver Labs Europe} \\
HSE University \\
\{stroshin, nchirkova\}@hse.ru}

\begin{document}
\maketitle
\begin{abstract}
Deep learning models are widely used for solving challenging code processing tasks, such as code generation or code summarization. Traditionally, a specific model architecture was carefully built to solve a particular code processing task. However, recently general pretrained models such as CodeBERT or CodeT5 have been shown to outperform task-specific models in many applications. While pretrained models are known to learn complex patterns from data, they may fail to understand some properties of source code. To test diverse aspects of code understanding, we introduce a set of diagnostic probing tasks. We show that pretrained models of code indeed contain information about code syntactic structure, the notions of identifiers, and namespaces, but they may fail to recognize more complex code properties such as semantic equivalence. We also investigate how probing results are affected by using code-specific pretraining objectives, varying the model size, or finetuning.
\end{abstract}
\section{Introduction}

Deep learning and especially Natural Language Processing (NLP) methods have been widely and successfully adopted to process source code. Example tasks include code generation \cite{allamanis15-code-gen,  codex2021} where the task is usually formulated as to produce a code of a function given the natural description; code translation \cite{staticalcodetrans2013, transcoder2020nips} where the model needs to translate from one programming language to another; and code summarization \cite{codsum'10, leclair2020codegnn} where the task is to produce natural language (NL) description for a given code snippet. Deep learning is also widely used in discriminative tasks, such as automated bug search and repair \cite{Hellendoorn2020Global}.

\begin{figure*}[t]
    \centering
     \includegraphics[width=1\linewidth]{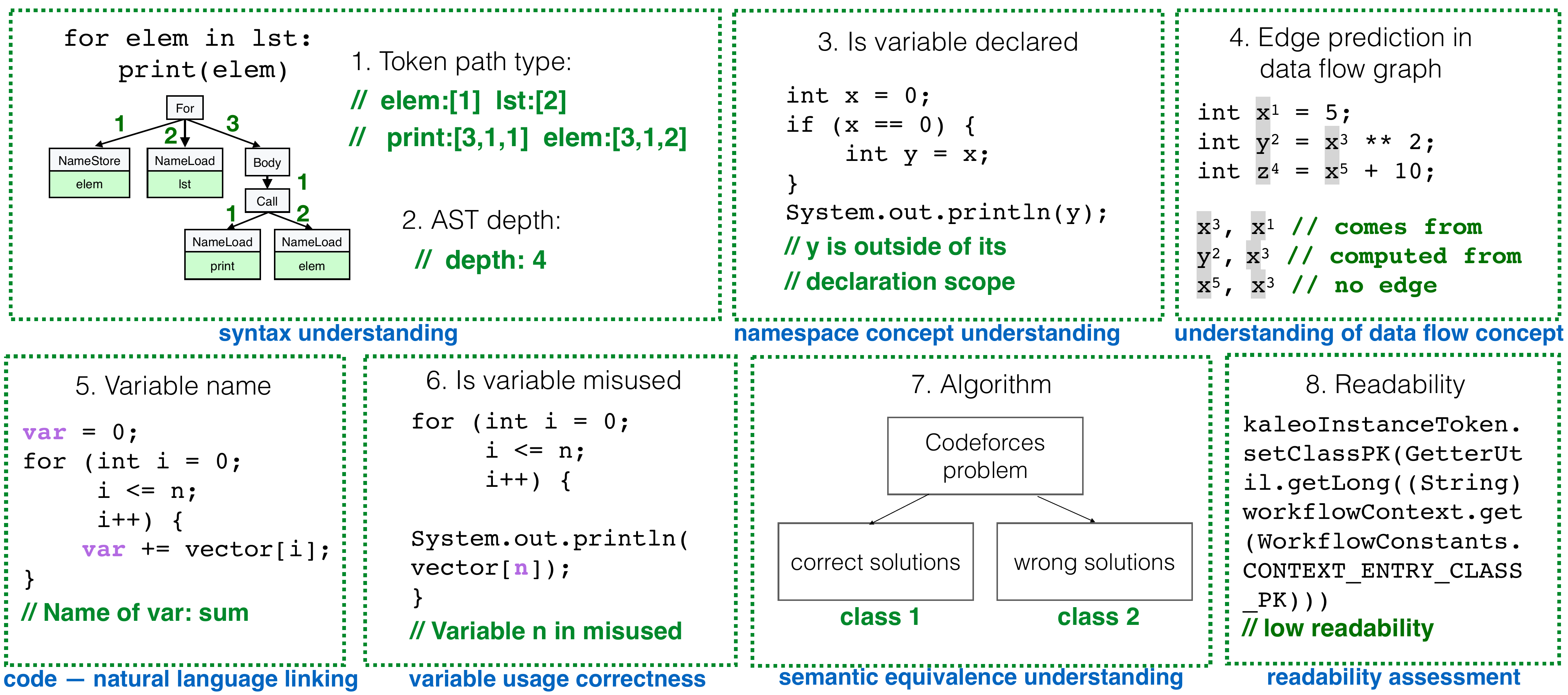}
    \caption{Illustration of the proposed probing tasks testing various aspects of code understanding.}
    \label{fig:tasksill}
\end{figure*}

In recent years, the focus has shifted from developing task-specific models incorporating prior knowledge about the task, to relying on general pretrained models of code such as CodeBERT~\cite{CodeBERT} or CodeT5~\cite{CodeT5}. These models, once pretrained, can be finetuned on the downstream tasks with a little additional cost, surpassing task-specific models. While the performance of the models is high on a wide range of downstream tasks~\cite{CodeXGLUE}, the boundary between what the models know and where they fail remains hidden behind the complexity of the downstream tasks. The lack of interpretability of pretrained models limits their practical use. At the same time, a deeper examination of model's understanding of source code may increase developers' trust and broaden the applicability of pretrained models. 

In NLP, there is an established probing approach for a more fine-grained examination of the knowledge of various aspects of the language, e.g. morphology, syntax, or discourse understanding \cite{belinkov-etal-2020-linguistic, context_probing, koto-etal-2021-discourse}. Probing usually means training a linear model on top of hidden representations of a model for various simple tasks, e.g. to predict a part-of-speech tag, to detect whether a sentence was corrupted, or to estimate the number of objects in the main clause \cite{conneau-etal-2018-cram}. 
Probing experiments may suggest ways to improve
the quality of the pretrained model or provide recommendations on how to tune the model better in applied tasks~\cite{probing_review}.

Inspired by the insights probing provided in NLP, we develop probing tasks to understand the extent to which the current state-of-the-art pretrained models capture structural and semantic properties of source code.
Our contributions are as follows:
\begin{itemize}
    \item we introduce a set of syntactic and semantic probing tasks, suitable for testing diverse aspects of code understanding;
    \item we study an effect of the model choice, pretraining objective choice, and model size
    on probing results;
    \item we use probings to highlight which information about  code is preserved by finetuned models in different downstream tasks.
\end{itemize}
We release our   \href{https://github.com/serjtroshin/probings4code}{code}\footnote{https://github.com/serjtroshin/probings4code}.


\section{Probing tasks}
\label{sec:tasks}
We probe pretrained models of code using linear regression or classification trained on top of code representations extracted from each layer of each model  (layers weights are not finetuned) \cite{DBLP:journals/corr/AlainB16}. 

We develop auxiliary tasks (with synthetic data or data borrowed from other works) that test models' understanding of various properties of source code: strict syntactic structure, the notions of data flow and namespaces, naming, semantic equivalence, and readability. We consider both \textit{global} tasks (predicting a property of the whole code snippet) and \textit{local} tasks (predicting a property of a particular token or a group of tokens). For each task, we introduce a simple but as strong as possible baseline. Figure~\ref{fig:tasksill} illustrates all tasks. For all classification tasks, we measure test error ($1-accuracy$), and for ``AST depth" regression task, we calculate the mean absolute error $MAE(y_{true}, y_{pred}) = \frac{1}{N}\sum_{i=1}^{N}|y^i_{true} - y^i_{pred}|$. 

\paragraph{Notation.}
Pretrained models of code usually follow the standard NLP methodology: representing a code snippet as a sequence of subtokens, e.~g. byte-pair encoding subtokens, and pretraining the model on a large corpora of source code using masked language modeling.
We denote the sequence of subtokens as $s_1, \ldots, s_m$. 
Let us denote $t(s_i)$ a mapping from a subtoken $s_i$ into a corresponding code token $t(s_i)$, e.g. for a subtoken sequence [\spverb|▁(|, \spverb|▁for|,  \spverb|▁public|, \spverb|▁get|, \spverb|Status|], $t(\spverb|▁get|) = \spverb|getStatus|$. For each subtoken $s_i$, we extract the model's embedding $\mathbf{w_i^\ell}\!\in\!R^{d}$ for a particular layer~$\ell$, where $d$   is the size of hidden representations.

\subsection{Token Path Type}
The first two probing tasks test whether pretrained models contain information about the syntactic structure of code.
The first task consists of predicting the position of a token in the Abstract Syntax Tree (AST). Given a subtoken $s_i$ and the corresponding embedding $\mathbf{w_i^\ell}$, the task is to predict the path type from the root to the $t(s_i)$ token, e.g. $[1, 2, 1]$, meaning go to the first child, then to the second one, then to the first one. This task, which is a local task, is formulated as a classification problem by selecting target subtokens corresponding to $15$ most frequent path types. As a baseline, we consider constant prediction w.~r.~t. a subtoken, i.~e. we select the most frequent class (path type) for each subtoken in the vocabulary.



\subsection{AST depth}
The second syntactic task is defined on a code snippet level (global task) and consists of predicting the depth of the AST built from the snippet (regression problem). The baseline for this task is defined as a linear regression trained on a single feature -- the number of tokens in the code snippet, this baseline outperforms the median depth baseline computed over the whole dataset. 


\subsection{Is Variable Declared}
This task tests the model's understanding of the notion of namespaces. The model is asked, whether there is an ``undeclared variable name" error for a particular expression with an identifier. For example, in the first code snippet the identifier $y$ is correctly used after a declaration:

\begin{verbatim}
int x = 0; 
if (x == 0) { 
    int y = x; // declare
    System.out.println(y); // use
}
\end{verbatim}
However, in the second snippet there is an error, since $y$ is outside of the scope of it's declaration:
\begin{verbatim}
int x = 0; 
if (x == 0) { int y = x; }
System.out.println(y); // y is undeclared
\end{verbatim}

We generate positive and negative examples using the following procedure. For a code snippet, we find a variable name declaration, e.g. \spverb|float x = 0|. Next, we find a random place in code after the variable declaration where we can insert a printing expression e.g. \spverb|System.out.println(x);|, and define a label for binary classification analyzing variable scopes: is variable declared before used?
The task is formulated for the mean subtoken embedding of the inserted variable name (local task). The baseline in this task is a constant prediction that the variable is declared.


\subsection{Edge Prediction in Data Flow Graph}
The next task measures to what extent a model encodes the information about the data flow. 
Given two tokens, the task is to predict a data flow edge between them. There can be no edge (negative example), a ``comes from" edge, or a ``computed from" edge. The task is formulated as classification of a pair of tokens (their mean embeddings over subtokens are concatenated), this is the local task. 



In addition to existent data flow edges we select a roughly equal number of ``no edge" examples by selecting random pairs of nodes from AST with suitable node types (e.g. pairs of identifiers, constants, etc.). As a baseline, we predict the most frequent edge type for the corresponding pair of tokens, which outperforms the most frequent class baseline.

\subsection{Variable Name}
The next task targets the ability to link code elements and their natural language descriptions. A model should predict a variable name, given a code snippet with all occurrences of the original name replaced with a placeholder \spverb|var|. This task requires semantic understanding of the variable's role in the program (local task).

We formulate this task as classification, targeting only $15$ most popular identifier names. The feature vector is a mean hidden representation for all occurrences of the identifier in code. In such way, the model should be able to predict the identifier name based on the context in which the variable was used. The baseline in this task is defined by the bag-of-words model: we count occurrences of all subtokens in the code snippet, convert them to tf-idf values and train a linear classifier on these features. This baseline substantially outperforms the constant baseline which always predicts the most frequent variable name.

\subsection{Is Variable Misused}
The next local task tests the ability of the model to detect the variable misuse bug \cite{Hellendoorn2020Global}. We introduce variable misuse by randomly assigning ``wrong" identifier name copying from another identifier from the same code snippet. We add ``correct" code snippet for each ``wrong" snippet, formulating the task as a binary classification problem, where the input is identifier's subtokens (mean embedding). The baseline is the bag of words predictor, which, for this task, is better than most frequent class predictor.

\subsection{Algorithm}
The next (global) semantic task also tests the ability of models to distinguish computationally equivalent codes from other codes. To obtain a dataset for this task, we select a simple problem from the CodeForces competition~\footnote{https://codeforces.com/problemset/problem/1671/A}, which can be reformulated as to check if each character in a string has a neighbor equal to it. We download ``wrong answer" and ``accepted" Python submissions from the contest and filter out too long codes ($>1000$ characters) obtaining $550$ ``accepted" and $384$ ``wrong answer" submissions.

The task is to distinguish ``correct" code from ``wrong" and formulated as a binary classification problem. The task requires deeper understanding of the data and control flow since the ``accepted" and ``wrong answer" solutions are usually very similar visually. It should be hard for a model to make predictions based only on spurious surface or syntactic features to succeed in this task.
As a baseline, we again use the bag-of-words model described above.


\subsection{Readability}
Finally, we consider a readability property of code. Readability defines how easy code is for the programmers to understand and maintain. Generally, readability depends on visual appearance of code (spaces, new lines etc), the meaningfulness of variable and function names, the quality of comments, and the particular algorithmic implementation (the same algorithm could be written in different ways, some of them more and some of them less readable). We use the 200 examples dataset provided by~\citet{readability} and obtained by collecting a set of functions and asking developers to rate readability on the scale from 1 to 5 (several ratings per example). The task is then converted by the authors to binary classification by treating all snippets with rating $\leqslant 3.6$ as not readable and the rest ones as readable, as in \citet{readability}.  This is a global task and as the baseline we use the bag-of-words model which outperforms the constant most frequent class prediction.

\section{Models}
\label{sec:models}
In this section, we briefly describe the models to be compared. We have selected several widely used pretrained models, which vary in the model architecture, pretraining objective, model size, and training datasets.

\subsection{CodeBERT}
CodeBERT \cite{CodeBERT} is one of the first attempts to pretrain a Transformer-based encoder model for source code representation learning and comprehension. It is a 12 layer encoder model based on RoBERTa-base (125M) \cite{roberta} and trained with masked language modeling and replaced token detection objectives. The model is trained on 6M CodeSeachNet dataset \cite{codesearchnet}, composed of functions from 6 programming languages (Java, Python, JavaScript, PHP, Ruby, Go) and NL comments.

\subsection{GraphCodeBERT}
GraphCodeBERT~\cite{guo2021graphcodebert} extends the work of \citep{CodeBERT}, by introducing data flow-related objectives. They encourage the models to learn structure-aware representations by predicting randomly selected ``comes from" data-flow edges.

\subsection{PLBART}

\citet{plbart} introduced a 140M parameter PLBART model with 6 encoder and 6 decoder layers. The model is based on the BART~\cite{bart} architecture. The authors released a PLBART~\footnote{\texttt{https://github.com/wasiahmad/PLBART}} checkpoint pretrained on the data collected by \citet{transcoder}, which is 470M Java, 210M Python functions/methods, and pretrained the 47M NL descriptions. They release a PLBART\_large checkpoint as well (400M, 12 layer encoder, 12 layer decoder).

\subsection{CodeT5}
 CodeT5 \cite{CodeT5} is an encoder-decoder model based on the T5 \cite{t5} architecture and pretrained on 8.35M functions in 8 programming languages (Python, Java, JavaScript, PHP, Ruby, Go, C, and C\#). The model combines the masked language modeling objective with code-specific objectives, including identifier tagging and predicting variable names. We experiment with two released model checkpoints~\footnote{\texttt{https://github.com/salesforce/CodeT5}}: CodeT5-base (220M) and CodeT5-small (60M).

\begin{figure*}[ht!]
    \centering
         \includegraphics[width=\linewidth]{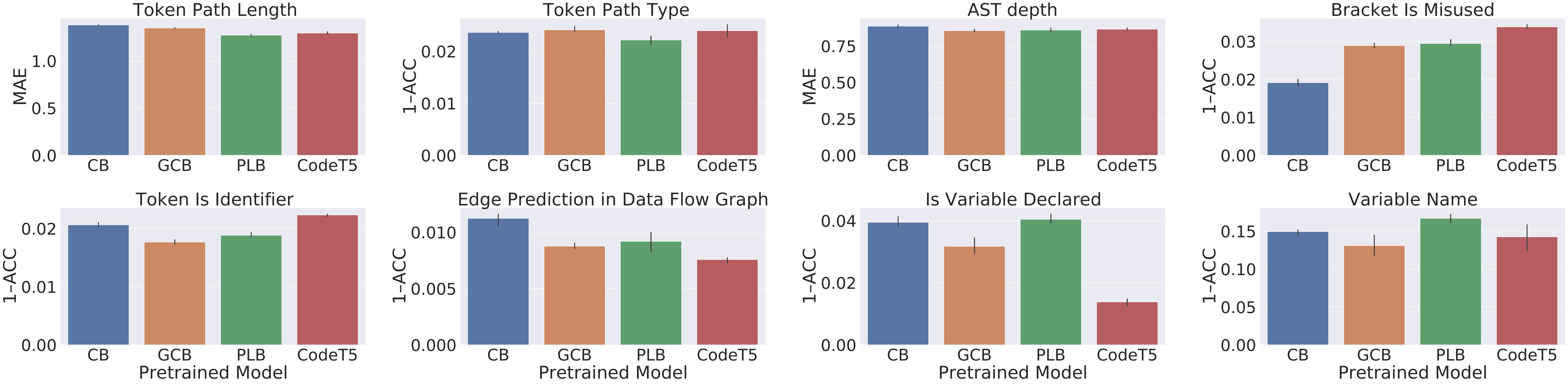}
        \caption{Results for the best performing layer representations, for all probing tasks. 
        Metrics are the lower the better. 
        }
        \label{fig:compare}
\end{figure*}

\subsection{CodeGPT2}
CodeGPT2 \cite{CodeXGLUE} is a decoder only model  based on GPT-2 architecture \cite{gpt2}. The 117M model consists of 12 layers and is pretrained on the  CodeSearchNet \cite{codesearchnet} dataset. We used \textit{CodeGPT-small-java-adaptedGPT2} checkpoint~\footnote{
\texttt{https://huggingface.co/microsoft/}
\,\,\,\texttt{/CodeGPT-small-java-adaptedGPT2}
}, that is initialized from GPT-2 model and then trained on code corpus.

\subsection{BERT}
We also consider the text-based model, BERT, to understand the effect of code-specific pretraining. We use a 110M 12-layer BERT model \cite{bert}, \textit{"bert-base-cased"} checkpoint from Hugging Face~\footnote{\texttt{https://huggingface.co/bert-base-cased}}, trained on the  Book Corpus \cite{moviebook}.

\section{Experimental setup}
\label{sec:setup}

\subsection{Data and preprocessing}
For our experiments which involve synthetic data (first 6 tasks), we use the test dataset provided by \citet{plbart} consisting of 10k examples of Java functions and methods with removed comments and new line symbols, and standardized code snippets. For the two remaining tasks the datasets were mentioned in the task descriptions.  For each pretrained model, we apply it's subtokenization procedure. All models have a limit of 512 input subtokens. We crop subtoken sequences that are longer than 512 subtokens.
We use commonly used open access datasets intended for research purposes.

\subsection{Probing details}
For each probing task, we average results over four runs using 4-fold cross-validation. 
For each model, we use a single checkpoint as usually only one checkpoint is released. 

Each pretrained model returns representations for a sequence of subtokens $s_1, \ldots, s_m$, e.~g.~from byte-pair encoding. When the task is formulated on a code snippet level, the layer-wise embeddings of the snippet are obtained by averaging subtoken embeddings, following~\cite{koto-etal-2021-discourse}.

For the probing models, we use linear models from scikit-learn (0.24.2) \cite{scikit-learn}, including
 \textit{SGDClassifier} with logistic regression loss for classification tasks (we select optimal alpha parameter via grid search over $[0.0001, 0.001, 0.01, 0.1, 1, 10, 100]$ range, and set tolerance to $0.0001$); and \textit{RidgeCV} for regression tasks (grid search for alpha over $[0.0001, 0.001, 0.01, 0.1, 1, 10, 100]$ range). In addition to linear probings, we also probe pretrained models with a 3-layer MLP (see  Appendix~\ref{sec:appendix_MLP}), however, the results for MLP are similar.
 

\section{Experiments}
Our research questions are as follows:
\begin{itemize}
    \item
    To what extent do the models pretrained on code capture information about source code properties? 
    \item Does multitask pretraining with code-specific objectives provide richer representations?
    \item How does the model size 
    affect probing results? Which representations are better: provided by the encoder or by the decoder? Which layers provide better representations?
    \item Does finetuning preserve syntactic and semantic information in different downstream tasks?
\end{itemize}

We used a single Tesla V100 GPU for the forward pass to collect embeddings, and 4 CPU for training linear models. Our total computational budget is 864 CPU hours and 20 GPU hours.

\label{sec:experiments}
\subsection{Comparison of different models}
\label{sec:pretrained}

In this subsection, we study the performance of different pretrained models in all probing tasks. 
In this experiment, we report the results for the best performing layer representation for each model: the layer is chosen using the first fold and the results are averaged over three remaining folds. Figure~\ref{fig:compare} presents the results.

Overall, we observe that the probing performance of pretrained models exceeds the performance of the simple baseline in all tasks. However, in the semantic-related  ``Readability" and ``Algorithm" tasks, the pretrained models are very close to the simple baseline and thus do not capture much more information relevant to these tasks. The BERT model pretrained on textual data performs worse than the models pretrained on code in all tasks except the semantic-related ``Readability" and ``Algorithm" tasks, where all pretrained models perform similarily. We conclude that \textit{models pretrained on code contain knowledge about basic source code properties but lack a deeper semantic understanding of code.}

Comparing different models, 
we find that the models pretrained with code-specific objectives, GraphCodeBert and CodeT5, are better or on par with other models for all tasks. 
In ``Edge Prediction in DFG", GraphCodeBERT performs best because it uses the edge prediction objective during pretraining. 
Similarly in ``Variable is Undeclared", ``Is Variable Misused" CodeT5 and GraphCodeBERT perform best potentially because they use the variable-related pretraining objectives. CodeGPT2 performs worse for ``Is Variable Misused" task, because it only sees the left context which may be not enough to predict the misused variable. 
To sum up, \textit{models pretrained with code-specific objectives, CodeT5 and GraphCodeBERT, show consistent gain for the tasks related to their pretraining objectives, over other models, pretrained with single objectives, or perform on par with them. } 


To better understand how pretrained models perform on each task, we perform an ablation study masking different code components: identifiers, keywords, and punctuation. This ablation study is described in Appendix~\ref{sec:appendix}. The main finding is that masking punctuation hurts the probing performance of the model pretrained on source code in almost all tasks, while masking language keywords and renaming identifiers do not have much effect (except the variable naming task where renaming identifiers has a significant effect).


\subsection{Encoder vs Decoder}
\label{sec:transfer}

This subsection compares the representations of the encoder and the decoder. We consider representations of two encoder-decoder models, PLBART-base and CodeT5-base. Table~\ref{table:plbart_codet5} compares best performing encoder representations and best performing decoder representations for all probing tasks. We observe that \textit{in almost all probing tasks, the decoder representations perform worse or on par with the encoder representations}. In some tasks, e.~g. ``Is Variable Misused", the decoder shows much worse results than the encoder. A possible explanation is that the aim of the encoder is to provide rich representations for the decoder, hence the encoder is more suitable for information extraction. 

\subsection{The effect of the model size}

In this subsection, we are interested whether larger models capture more information about the source code properties than smaller models. 
Table~\ref{table:plbart_codet5} reports the performance of CodeT5-base and CodeT5-small models, and of PLBART-large and PLBART-base models (other models are not available in variable sizes). 
We find that \textit{in three variable related tasks the larger models expectantly perform better than smaller models but in the majority of the tasks the performance is similar}.

\subsection{Per-layer probing performance}
\label{sec:layers}

\begin{figure}[!ht]
    \centering
         \includegraphics[width=0.9\linewidth]{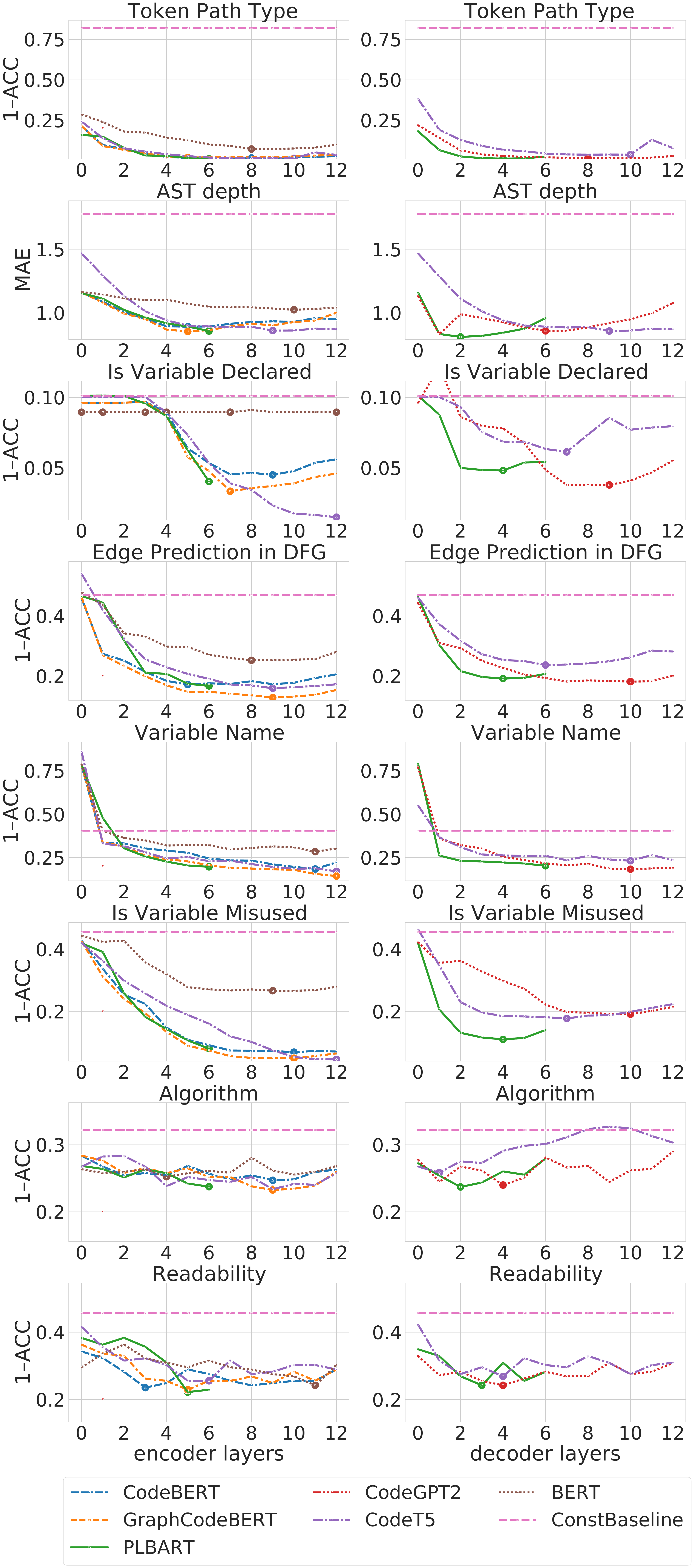}
        \caption{Per-layer probing performance of four pretrained models. Dots highlight the best layers for a particular model.}
        \label{fig:which_layers_are_the_best}
\end{figure}

We now analyse probing results for different Transformer layers. 
Figure~\ref{fig:which_layers_are_the_best} shows the per-layer performance of all considered pretrained models.
In syntax-related, namespace-related, data flow-related, algorithm-related and readability-related tasks, \textit{middle layers (4--10) usually provide the most informative representations}. 
In the ``Variable Name" (and partly in ``Is Variable Misused"), the last layers consistently perform better because the task is closely related to the masked language modeling objective which is usually solved on top of last layers. 

\subsection{The effect of finetuning}

In this section, we study the effect of finetuning on probing results. Specifically, we are interested 1) whether finetuned models preserve information contained in pretrained models; 2) does pretraining enrich the representations of finetuned models, compared with the representations of models trained from scratch.

In this section, we focus on the PLBART model and finetune it for 5 downstream tasks: 3 generative tasks (Code Translation from Python to Java, Java Code Generation based on natural language descriptions, Java Code Summarization into textual description) and 2 discriminative tasks (Clone Detection, Defect Prediction).
We use the AVATAR dataset~\cite{ahmad-etal-2021-avatar} in the Code Translation task and CodeXGLEU benchmark \cite{CodeXGLUE} in other tasks (MIT license). We use scripts for PLBART finetuning on these tasks provided in PLBART~\footnote{\texttt{https://github.com/wasiahmad/PLBART}} and AVATAR~\footnote{\texttt{https://github.com/wasiahmad/AVATAR}} repositories.

\begin{figure*}[t!]
    \centering
         \includegraphics[width=\linewidth]{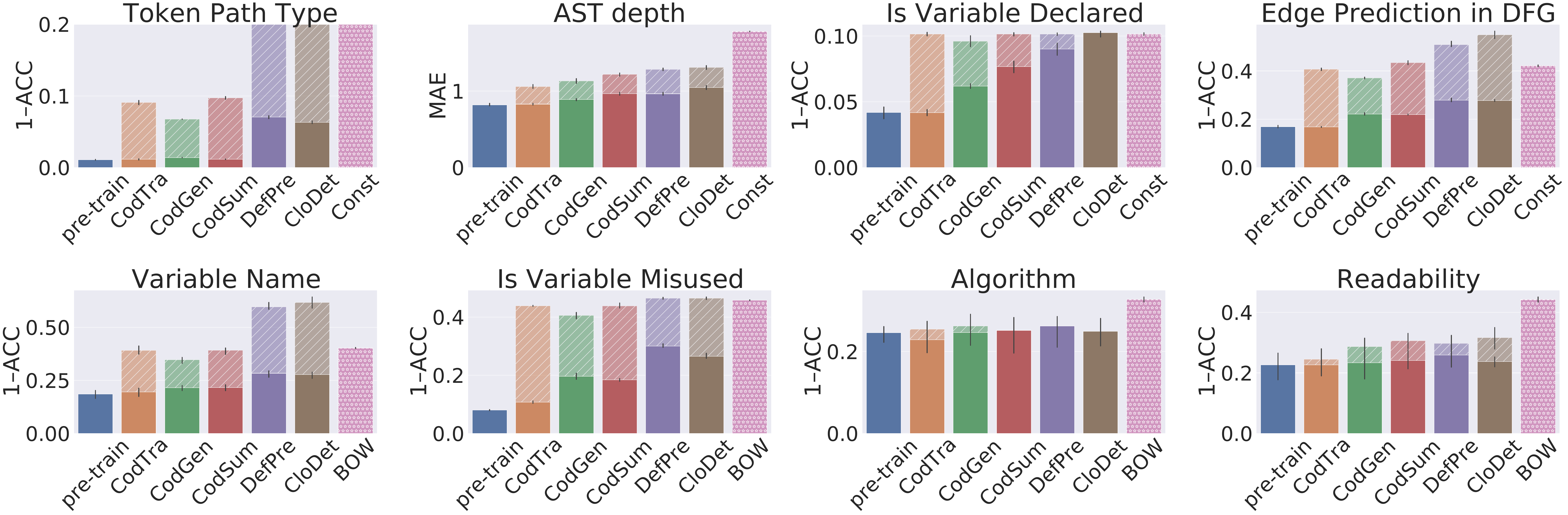}
        \caption{Results on the effect of finetuning. Pre-train (leftmost bar): pretrained-only checkpoint. The following bars: dark -- finetuned models, semi-transparent -- models trained from scratch. Results for 5 downstream tasks: Code Translation, Code Generation, Code Summarization, Defect Prediction, Clone Detection.}
        \label{fig:finetune}
\end{figure*}

Figure \ref{fig:finetune} compares 3 scenarios: the PLBART checkpoint after pretraining (leftmost bar),  checkpoints after PLBART finetuning on each of 5 downstream tasks (dark bars), and checkpoints after training from scratch on each of 5 downstream tasks (semi-transparent bars). We also include baselines for reference.

\textit{Models finetuned for discriminative tasks exhibit the highest information loss between the initial pretrained stage and the finetuned stage}, which may indicate that models trained on these tasks rely on some spurious features, rather than on code syntax or semantics.

\begin{table*}[!t]
\small
\centering
\begin{tabular}{l|ll|ll||ll|ll}

                                     &   \multicolumn{2}{c}{PLBART}       &    \multicolumn{2}{c}{CodeT5}    & \multicolumn{2}{c}{PLBART}  & \multicolumn{2}{c}{CodeT5}   \\
 Task                                &   encoder         & decoder        & encoder         & decoder      &  base & large  &    small & base                 \\ \hline
 
Token Path Type                      &   \textbf{0.011} &           0.012 &  \textbf{0.013} &           0.036 &  \textbf{0.011} &           0.014 &  \textbf{0.012} &           0.013 \\
AST depth                            &            0.866 &  \textbf{0.820} &  \textbf{0.867} &  \textbf{0.864} &  \textbf{0.820} &  \textbf{0.850} &  \textbf{0.863} &  \textbf{0.864} \\
Is Variable Declared                 &   \textbf{0.042} &  \textbf{0.049} &  \textbf{0.014} &           0.061 &           0.042 &  \textbf{0.019} &           0.025 &  \textbf{0.014} \\
Edge Prediction in DFG               &   \textbf{0.167} &           0.191 &  \textbf{0.161} &           0.230 &  \textbf{0.167} &  \textbf{0.167} &  \textbf{0.162} &  \textbf{0.161} \\
Variable Name                        &   \textbf{0.186} &  \textbf{0.194} &  \textbf{0.162} &           0.211 &  \textbf{0.186} &  \textbf{0.165} &           0.208 &  \textbf{0.162} \\
Is Variable Misused                  &   \textbf{0.080} &           0.112 &  \textbf{0.046} &           0.176 &           0.080 &  \textbf{0.053} &           0.064 &  \textbf{0.046} \\
Algorithm                            &   \textbf{0.239} &  \textbf{0.251} &  \textbf{0.228} &  \textbf{0.268} &  \textbf{0.246} &  \textbf{0.225} &  \textbf{0.235} &  \textbf{0.228} \\
Readability                          &   \textbf{0.226} &  \textbf{0.237} &  \textbf{0.247} &  \textbf{0.246} &  \textbf{0.216} &  \textbf{0.238} &  \textbf{0.242} &  \textbf{0.221} \\

\end{tabular}
\caption{Encoder vs decoder performance for PLBART-base and CodeT5-base; and comparison of small vs large models: PLBART-base vs PLBART-large, and CodeT5-small vs CodeT5-base. Metrics: MAE for ``AST depth", otherwise test error (1-accuracy).}
\label{table:plbart_codet5}
\end{table*}

\textit{Among generative tasks, the Code Translation model exhibits almost no gap between pretrained and finetuned stages.} This could be attributed to having code as both input and output of the task. Code Generation and Code Summarization models have code only as either the input or the output of the task, and usually exhibit a slightly larger gap. 

As for models, trained from scratch for downstream tasks (semi-transparent bars), the overall trend is similar across the downstream tasks, but the absolute results are usually much worse, compared to finetuned models, and sometimes are close to simple baselines. \textit{The downstream tasks alone do not provide high-quality code representations.}

\section{Related Work}
\label{sec:related}

Probing became a universal tool in NLP for testing pretrained models' understanding or knowledge of various language aspects. A simple linear probing was used in \cite{gupta} to test whether referential knowledge is already encoded in word embeddings, while \citet{kohn} got insights into the behaviour of word embedding in terms of morphological and syntactical properties. Probing tasks were developed to evaluate sentence embeddings \cite{ettinger, conneau} whether they incorporate compositional, or surface (length of the sentence), syntax (tree depth, top constituent), and text semantics (e.g. tense of a sentence) knowledge. \citet{hewitt-manning-2019-structural} proposed more complex probing tasks, questioning the possibility to parse the whole dependency trees from the sentence embeddings using a metric learning approach. More recent studies for language models include the study of emerging capabilities of large language models \cite{emerging_abilities}.  We refer to~\citet{probing_review} for a broad review of existing probing works in NLP.


In the context of source code, \citet{code_prob} made the first steps towards probing pretrained models. However, they only consider four simple tasks and tree code models, CodeBERT, CodeBERTa and GraphCodeBERT. In contrast to their work, we propose a wider set of tasks, including several token-wise tasks, consider a wider range of pretrained models, and investigate various dimensions, including different pretraining objectives, model sizes, and the effect of finetuning. 

A line of work investigate pretrained models for code in different directions. \citet{code_search_anon} show that CodeBERT relies on the high token overlap between query and code solving code search task rather than deeper syntax or semantic features, and \citet{code_attention} shows that BERT trained on code pays more attention to identifiers and separators. Our work provides another view on the analysis of pretrained models of code, from the probing perspective, and complements these results.

Recently created BIG-bench benchmark \cite{bigbench} contains a number of  challenging code-related probing tasks for testing large language models  capabilites, including programming synthesis and code summarization tasks, which are close to complex downstream tasks. In contrast, we aim at developing simple probing tasks targeting specific code understanding aspects.

\section{Conclusion and discussion}
\label{sec:c}
We presented a diagnosis tool, based on probing tasks, that can be used to estimate to which extent deep learning models capture the information about various properties of source code in their hidden representations. Our results show that pretrained models of code do contain information about code syntactic structure, the notion of namespaces, data flow, code readability and natural language-based naming. However, pretrained models show limited understanding of code semantics, which means that their usefullness in applied tasks requiring semantic understanding of code may be limited.  

Using code-specific pretraining objectives (CodeT5, GraphCodeBert) enriches the understanding of the code aspects addressed in the corresponding objective. This result may suggest practitioners to choose pretrained models which pretraining objectives are better aligned with the considered applied task.

We also found that finetuning may deteriorate the model's understanding of code properties, especially in classification downstream tasks. This may suggest including code-specific objectives in finetuning, especially if multi-stage finetuning~\cite{multistagef} is used.

\section*{Limitations}
\label{sec:limitations}
In this section, we discuss the limitations of our probing toolkit.

Our probing setup does not cover all possible aspects of source code. However, we were aiming at covering diverse properties of code.

Our experiments are limited to the two most popular high-level languages, which are usually used to evaluate pretrained model for code: Java (7 tasks) and Python (1 task). It would be interesting to compare the models on low-level languages like C/C++. 

The linear models used for probings may appear limited in their capacity, however, they were successfully used in a lot of NLP probing approaches~\cite{probing_review} and are well suitable for particular research questions considered in the paper. Moreover, we also experiment with a 3-layer MLP and find that our main results hold for MLP.

Finally, in this work, we only considered open-sourced pretrained checkpoints. It would be interesting to compare the performance of pretrained models across a wide range of model sizes.


\section*{Ethics Statement}
\label{sec:broader}
The main goal of this paper is to provide an empirical study of the existent models. Since we do not propose new models, there are no potential social risks to the best of our knowledge. Our work may benefit the research community providing more introspection to the current state-of-the-art models.

\section*{Acknowledgments}
The results were supported by the Russian Science Foundation grant \textnumero 19-71-30020.
The research was supported in part through the computational resources of HPC facilities at NRU HSE.




\bibliography{anthology,custom}
\bibliographystyle{acl_natbib}




\clearpage

\appendix
\section{Ablation study}
\label{sec:appendix}

In this section we perform an ablation study of different code components to understand which component of code is important for each of the probing tasks, i. e. identifier names, language specific keywords (e.g. ``for", ``if"), or punctuation (e.g. ``(", ``:", ``<") . We experiments with two pretrained models: pretrained on code (PLBART) and text model (BERT). For ablation of identifiers, we rely on the methodology of \cite{chirkova_empirical} and apply syntax-preserving anonymization, replacing identifiers inside a code snippet with placeholders (``var1", ``var2", ``var3"). To ablate language-specific keywords or punctuation, we simply replace them with ``MASK".

\begin{figure*}[t!]
     \includegraphics[width=\linewidth]{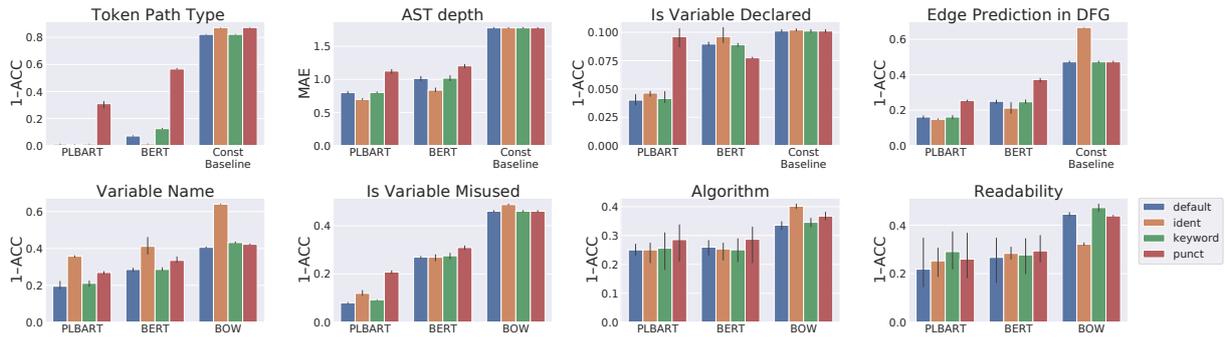}
    \caption{The results for ablation study: no ablation (default), anonymization of identifiers (ident), masking keywords (keyword), and masking punctuation (punct). Metrics are the lower the better.}
    \label{fig:ablation}
\end{figure*}

The ablation results are presented in Figure~\ref{fig:ablation}. Overall, for all tasks the most influential component is punctuation: masking it hurts the quality the most, except for the ``Variable Name" task, where anonymizing identifier leads to the worst quality. With punctuation being masked, PLBART model is close to the quality of the text BERT model or performs even worse.


In contrast, masking language-specific keywords does not hurt the performance significantly. 

To conclude, the models pretrained on code rely heavily on punctuation, for almost all tasks, and also rely on identifier names for variable related tasks.


\section{MLP}
\label{sec:appendix_MLP}

Linear probings may appear limited, thus we also include the results for 3-layer MLP model for comparison. We implement an MLP in PyTorch \cite{PyTorch} with ReLU nonlinearity and hidden size $128$. We train it with AdamW \cite{AdamW} optimizer with batch size $512$ and we use Optuna \cite{optuna} with 10 trials to search over learning rate (high=$0.1$, low=$0.0001$, $log$ domain) and weight decay (high=$0.1$, low=$0.00001$, $log$ domain) minimizing error on validation set ($0.1$\% of train set) for regression/classification tasks. We reduce learning rate by factor $0.1$ with patience $5$, use early stopping with patience $10$, and maximum number of update $5000$.

\begin{figure*}[t!]
    \centering
         \includegraphics[width=\linewidth]{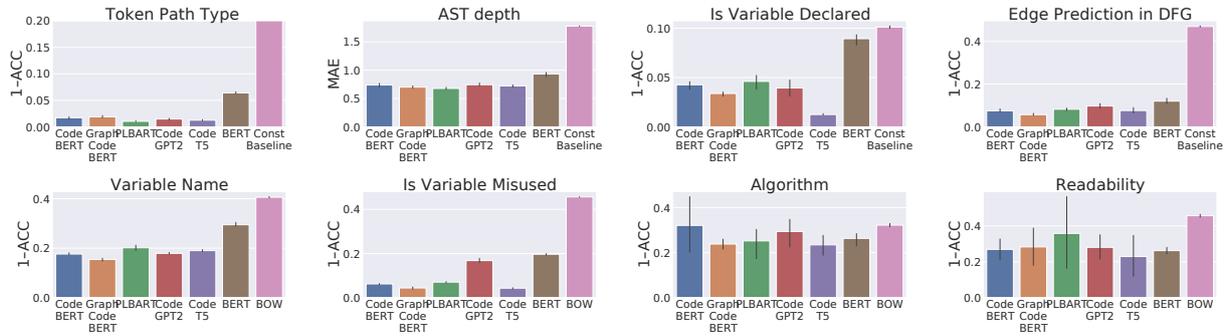}
        \caption{Results for the best performing layer representations, for all probing tasks, \emph{3-layer MLP}. 
        Metrics are the lower the better. 
        }
        \label{fig:compare_MLP}
\end{figure*}

\begin{figure*}[t!]
    \centering
         \includegraphics[width=\linewidth]{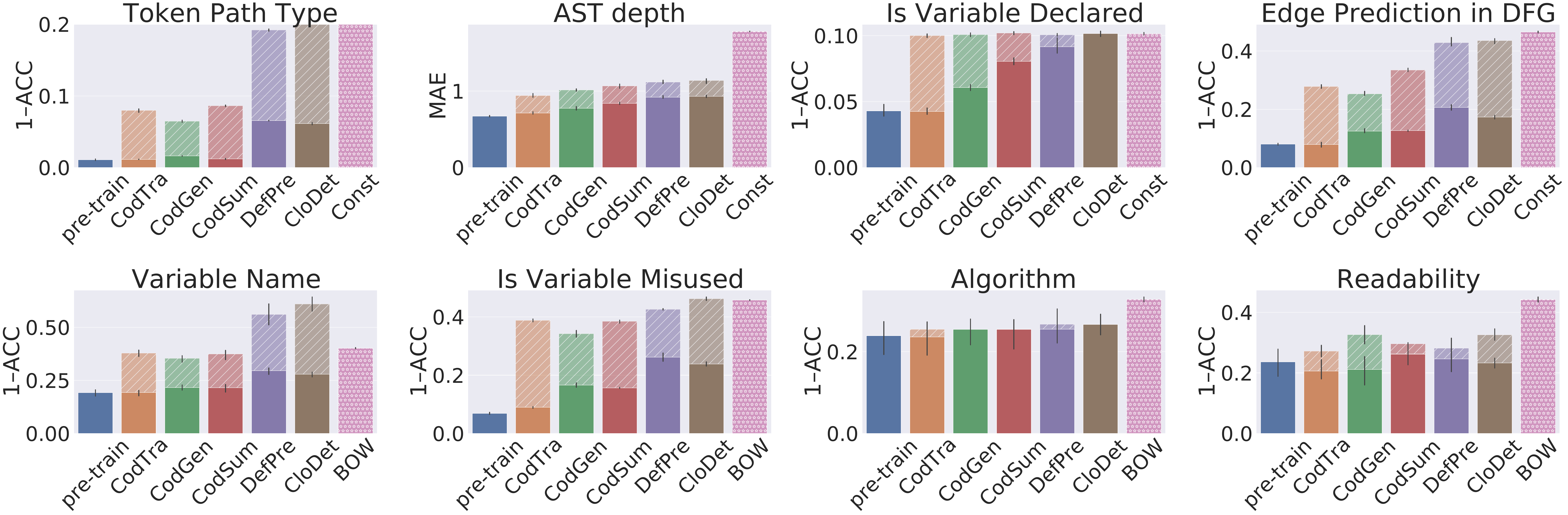}
        \caption{Results on the effect of finetuning, \emph{3-layer MLP}. Pre-train (leftmost bar): pretrained-only checkpoint. The following bars: dark -- finetuned models, semi-transparent -- models trained from scratch. Results for 5 downstream tasks: Code Translation, Code Generation, Code Summarization, Defect Prediction, Clone Detection. }
        \label{fig:finetune_MLP}
\end{figure*}

The results for MLP (Figures~\ref{fig:compare_MLP},\ref{fig:finetune_MLP})  are very similar to the results with linear models, both quantitatively and qualitatively. For edge prediction in data flow graph, MLP outperformes linear model significantly, but for other tasks the results are roughly the same.




\end{document}